\documentclass[prl,aps,amsfonts,amsmath,amssymb,twocolumn,superscriptaddress]{revtex4-2}

\usepackage{graphicx}
\usepackage{bm}
\usepackage{hyperref}
\usepackage{IEEEtrantools}
\usepackage{natbib}
\usepackage{float}
\usepackage{siunitx}
\usepackage{xcolor}

\restylefloat{table}

\newcommand{\tsi}{(TaSe$_4$)$_2$I }

\begin{document}


\title{Kramers-Weyl fermions in the chiral charge density wave material (TaSe$_4$)$_2$I}

\author{Soyeun Kim}
\affiliation{Department of Physics, University of Illinois at Urbana-Champaign, Urbana, 61801 IL, USA}
\affiliation{Materials Research Laboratory, University of Illinois at Urbana-Champaign, Urbana, 61801 IL, USA}

\author{Robert C. McKay}
\affiliation{Department of Physics and Institute for Condensed Matter Theory,
University of Illinois at Urbana-Champaign, Urbana, IL, 61801-3080, USA}

\author{Nina Bielinski}
\affiliation{Department of Physics, University of Illinois at Urbana-Champaign, Urbana, 61801 IL, USA}
\affiliation{Materials Research Laboratory, University of Illinois at Urbana-Champaign, Urbana, 61801 IL, USA}

\author{Chengxi Zhao}
\affiliation{Department of Materials Science and Engineering, University of Illinois at Urbana-Champaign, Urbana, IL 61801, USA}
\affiliation{Materials Research Laboratory, University of Illinois at Urbana-Champaign, Urbana, 61801 IL, USA}

\author{Meng-Kai Lin}
\affiliation{Department of Physics, University of Illinois at Urbana-Champaign, Urbana, 61801 IL, USA}
\affiliation{Materials Research Laboratory, University of Illinois at Urbana-Champaign, Urbana, 61801 IL, USA}
\affiliation{Department of Physics, National Central University, Taoyuan 32001, Taiwan}

\author{Joseph A. Hlevyack}
\affiliation{Department of Physics, University of Illinois at Urbana-Champaign, Urbana, 61801 IL, USA}
\affiliation{Materials Research Laboratory, University of Illinois at Urbana-Champaign, Urbana, 61801 IL, USA}

\author{Xuefei Guo}
\affiliation{Department of Physics, University of Illinois at Urbana-Champaign, Urbana, 61801 IL, USA}
\affiliation{Materials Research Laboratory, University of Illinois at Urbana-Champaign, Urbana, 61801 IL, USA}

\author{Sung-Kwan Mo}
\affiliation{Advanced Light Source, Lawrence Berkeley National Laboratory, Berkeley, California 94720, USA}

\author{Peter Abbamonte}
\affiliation{Department of Physics, University of Illinois at Urbana-Champaign, Urbana, 61801 IL, USA}
\affiliation{Materials Research Laboratory, University of Illinois at Urbana-Champaign, Urbana, 61801 IL, USA}

\author{T.-C. Chiang}
\affiliation{Department of Physics, University of Illinois at Urbana-Champaign, Urbana, 61801 IL, USA}
\affiliation{Materials Research Laboratory, University of Illinois at Urbana-Champaign, Urbana, 61801 IL, USA}

\author{Andr\'e Schleife}
\affiliation{Department of Materials Science and Engineering, University of Illinois at Urbana-Champaign, Urbana, IL 61801, USA}
\affiliation{Materials Research Laboratory, University of Illinois at Urbana-Champaign, Urbana, 61801 IL, USA}
\affiliation{National Center for Supercomputing Applications, University of Illinois at Urbana-Champaign, Urbana, IL 61801, USA}

\author{Daniel P. Shoemaker}
\affiliation{Department of Materials Science and Engineering, University of Illinois at Urbana-Champaign, Urbana, IL 61801, USA}
\affiliation{Materials Research Laboratory, University of Illinois at Urbana-Champaign, Urbana, 61801 IL, USA}

\author{Barry Bradlyn}
\email{bbradlyn@illinois.edu}
\affiliation{Department of Physics and Institute for Condensed Matter Theory,
University of Illinois at Urbana-Champaign, Urbana, IL, 61801-3080, USA}

\author{Fahad Mahmood} 
\email{fahad@illinois.edu}
\affiliation{Department of Physics, University of Illinois at Urbana-Champaign, Urbana, 61801 IL, USA}
\affiliation{Materials Research Laboratory, University of Illinois at Urbana-Champaign, Urbana, 61801 IL, USA}


\maketitle

\textbf{
The quasi-one-dimensional chiral charge density wave (CDW) material (TaSe$_4$)$_2$I has been recently predicted to host Kramers-Weyl (KW) fermions which should exist in the vicinity of high symmetry points in the Brillouin zone in chiral materials with strong spin-orbit coupling. However, direct spectroscopic evidence of KW fermions is limited. Here we use helicity-dependent laser-based angle resolved photoemission spectroscopy (ARPES) in conjunction with tight-binding and first-principles calculations to identify KW fermions in (TaSe$_4$)$_2$I. We find that topological and symmetry considerations place distinct constraints on the (pseudo-) spin texture and the observed spectra around a KW node. We further reveal an interplay between the spin texture around the chiral KW node and the onset of CDW order in (TaSe$_4$)$_2$I. Our findings highlight the unique topological nature of (TaSe$_4$)$_2$I and provide a pathway for identifying KW fermions in other chiral materials.}

\bigskip

The past decade has seen the prediction and discovery of a large number of topological materials, including 3D topological insulators\cite{fu2007topologicala,xia2009observation}, Dirac\cite{liu2014discovery,liu2014stable} and Weyl semimetals\cite{lv2015experimental,lv2015observation,xu2015discovery,xu2015discoverya}, and multifold chiral semimetals\cite{bradlyn2016dirac,Chang2017,sanchez2019topological,rao2019observation,schroter2019chiral,schroter2020observation}. These materials are characterized by nontrivial topology in the electronic band structure. They are typically identified using a combination of ab-initio calculations and symmetry-based methods that help to isolate unique spectroscopic features (such as Dirac and Weyl fermions) at high symmetry points and lines in the Brillouin zone (BZ). In many cases, these topological fermions arise due to the presence of certain crystalline symmetries\cite{vergniory2019high,tang2019efficient,zhang2019catalogue,Armitage2018,wieder2021topological}.

\begin{figure*}[bt!]
\includegraphics[width=2\columnwidth]{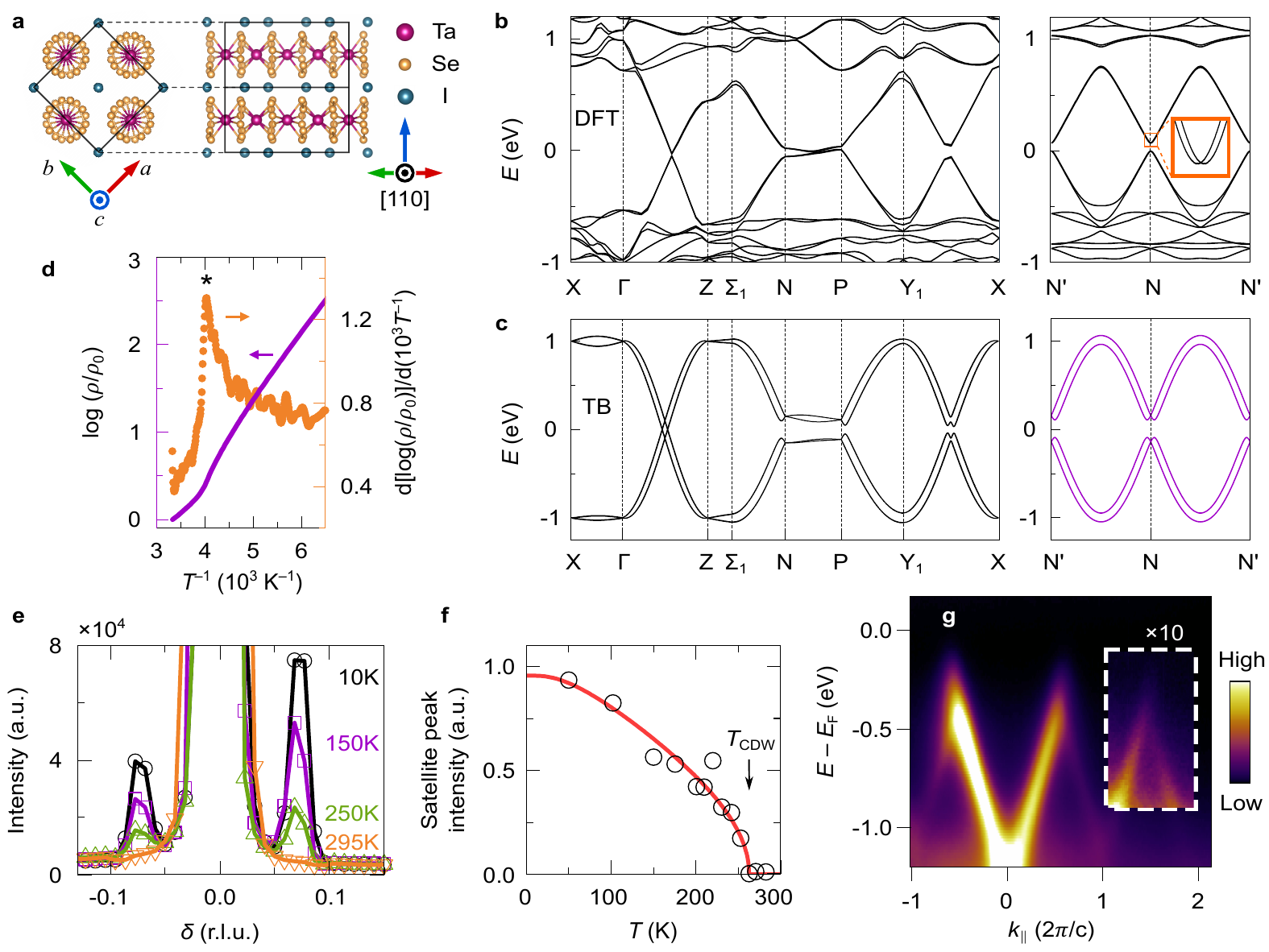}
\caption{\label{Fig_Characterization} \textbf{Crystal structure, experimental characterization, and Kramers-Weyl band structure of (TaSe$_4$)$_2$I.} \textbf{(a)} Crystal structure as seen from the (001) and (110) directions. Here (001) is along the Ta chain direction and (110) is the natural cleavage plane. A (conventional) unit cell is indicated by solid lines. \textbf{(b-c)} Electronic band structure of (TaSe$_4$)$_2$I as calculated using \textbf{(b)} density functional theory (DFT) and \textbf{(c)} a tight-binding model (TB) described in the SI \cite{Suppl}. The Kramers-Weyl crossing along the N'-N-N' path is shown in the right panels. \textbf{(d)} The logarithmic resistivity normalized by $\rho_0=\rho$(300 K). \textbf{(e-f)} Temperature dependent X-ray diffraction (XRD) results. \textbf{(e)} Line cuts of CDW satellite peaks for the Bragg reflection at ($h, k, l$) = ($-1$, 3, 4). \textbf{(f)} Temperature dependence of the CDW satellite peaks (symbols) as obtained from XRD data. Solid line is a fit with a BCS gap. $T_\text{CDW} \sim$ 260 K. \textbf{(g)} Synchrotron ARPES scan taken parallel to the $\Gamma Z$ (chain) direction at room temperature using 50 eV photon energy. The intensity in the dashed box is rescaled by 10.}
\end{figure*}

Recently, however, it has been realized that chiral crystals, characterized by the absence of orientation-reversing symmetries, with strong spin-orbit coupling (SOC) universally host topological Kramers-Weyl (KW) fermions\cite{Hasan2018}. In these materials, each time-reversal invariant momentum (TRIM) point in the BZ must feature a topologically charged Weyl point (or multifold fermion, if there are additional crystal symmetries) and is referred to as a KW point. Electronic bands near the KW point are split by SOC in all directions, forming pockets with opposite Chern number and an approximately radial (monopole-like) spin texture. Other novel features of these KW fermions include the circular photogalvanic effect \cite{Chang2017,flicker2018chiral, Rees2020,dejuan2017quantized,ni2020linear} and a longitudinal magneto-electric response \cite{Zhang2017, Wan2018,ni2021giant}.

As highlighted in Ref.\cite{Hasan2018}, a number of chiral crystals are predicted to host KW fermions, yet experimental confirmation remains elusive. This is primarily because the KW nodes are often far from a material's Fermi level and can also be in close proximity to other symmetry-enforced Weyl nodes. For example, Ref.~\cite{Hasan2018} predicted that the chiral charge density wave (CDW) compound (TaSe$_4$)$_2$I \cite{wang1983charge,maki_charge_1983} hosts KW nodes at its N TRIM point but at an energy above its Fermi level. (TaSe$_4$)$_2$I has also gained extensive interest recently as a Weyl-CDW candidate whereby its Fermi surface Weyl points (FSWP) are gapped by the onset of CDW correlations below the CDW transition temperature $T_\text{CDW}$\cite{Wujun2021}. This can lead to novel magneto-electric responses in the presence of an external magnetic field \cite{Gooth2019, Wang2013}. Whether the predicted KW nodes in (TaSe$_4$)$_2$I are affected by the CDW order remains an open question and could shed light into the interplay between strong correlations and Kramers-Weyl physics.

Here we use helicity-dependent laser ARPES, in combination with a tight-binding model and first-principles calculations to observe distinctive signatures of KW fermions in the chiral CDW compound (TaSe$_4$)$_2$I. Note that the surface of (TaSe$_4$)$_2$I is intrinsically n-doped due to iodine vacancies, which raise its chemical potential as observed in previous photoemission experiments \cite{Yi2021}. We take advantage of this to isolate conduction band features around the N TRIM point. We find that the helicity-dependent photoemission intensity is directly correlated with the unique (pseudo-)spin texture around this TRIM point, which confirms the presence of KW fermions in (TaSe$_4$)$_2$I. We further discover a decrease in the circular dichroic ARPES intensity with the onset of CDW order, suggesting a mixing of the KW chirality due to CDW correlations in addition to the previously identified mixing of the FSWPs.

\begin{figure*}[ht!]
\includegraphics[width=2\columnwidth]{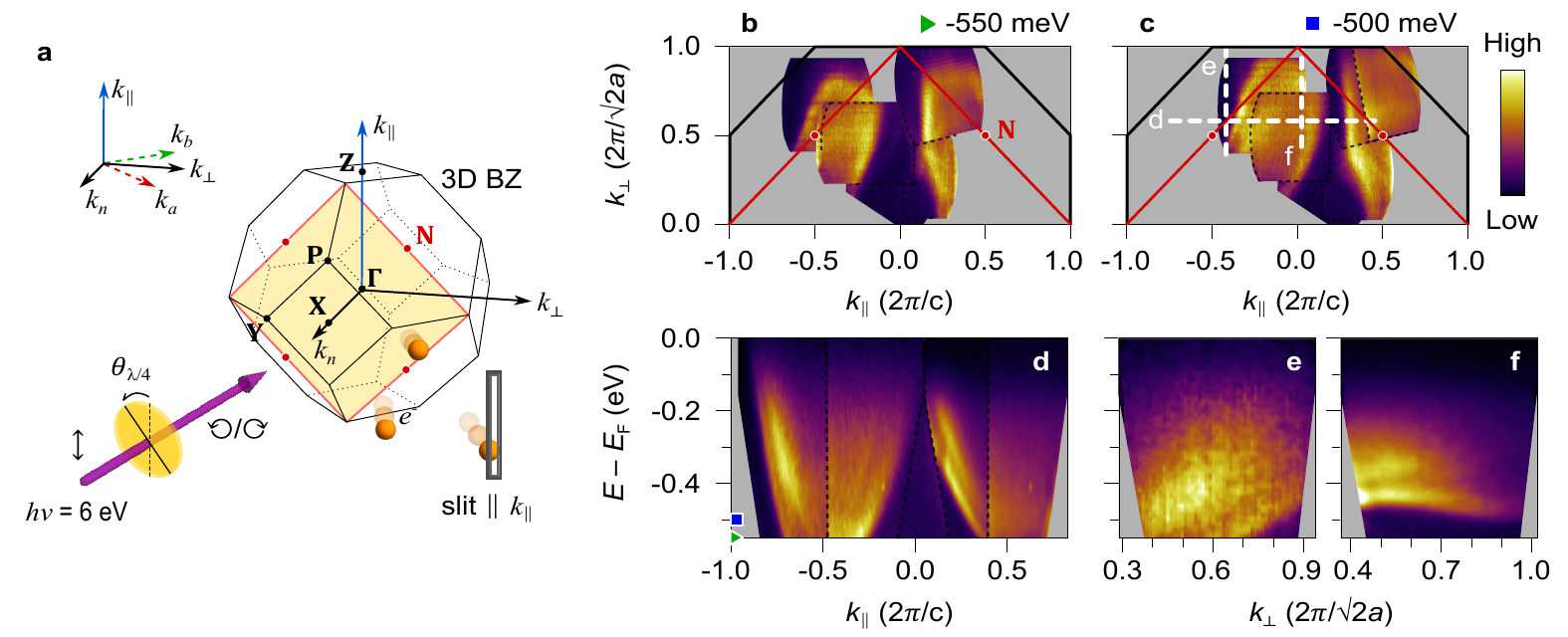}
\caption{\textbf{Laser-ARPES measurements on (TaSe$_4$)$_2$I using a photon energy of $h\nu = $ 6 eV taken at 280 K.} \textbf{(a)} Measurement geometry in relation to the 3D Brillouin zone (BZ). $k_{a}$/$k_{b}$ is the crystal $a/b$ direction. $k_\parallel$/$k_{\perp,n}$ is parallel/perpendicular to the Ta chain direction. The (110) cleavage plane corresponding to $k_n$ = 0.5 $(2\pi/\sqrt2a)$ is highlighted as the shaded plane. The light helicity is adjusted by rotating the fast axis of the $\lambda$/4 quarter waveplate ($\theta_{\lambda/4}$). \textbf{(b-c)} Constant energy maps taken at (b) $E-E_F$ = $-550$ meV and at (c) $E-E_F$ = $-500$ meV, summed over 20 meV. \textbf{(d-f)} Energy-momentum cuts along the white lines shown in (c).}\label{Fig_TSI_bigmap}
\end{figure*}

(TaSe$_4$)$_2$I has been studied extensively as a model quasi-one-dimensional system undergoing a CDW Peierls transition \cite{Gruner1988, Voit2000, TournierColletta2013}. As shown in Fig.\ref{Fig_Characterization}a, (TaSe$_4$)$_2$I consists of chains of Ta atoms surrounded by Se atoms along the $c$-axis. The chains are bonded weakly by I atoms, forming a needle-like crystal that naturally cleaves along the (110) plane. In Fig.~\ref{Fig_Characterization}b we show the band structure of (TaSe$_4$)$_2$I, computed using density functional theory (DFT). The left panel shows the bands along high-symmetry lines, and is consistent with known literature \cite{Hasan2018,Wujun2021,vergniory2019high}. In the right panel, we show the bands along the experimentally relevant path $(k_x,k_y,k_z) = (\pi/a, 0, k_z)$, with the N point at $k_z=\pi/c$, and the N' point at $k_z=-\pi/c$. We see in the inset that there is a small spin-orbit splitting visible, exposing a Kramers-Weyl fermion at the N TRIM point. To facilitate further theoretical calculations of photoemission intensity, we use this DFT input to construct a symmetry-inspired four-band tight-binding model. We use techniques from topological quantum chemistry \cite{bradlyn2017topological} to ensure this model reproduces the symmetry properties of the four bands closest to the Fermi level, as determined in the Topological Materials Database \cite{vergniory2021all}. Details of the model can be found in the SI \cite{Suppl}. The spectrum of the tight-binding model is shown in Fig.~\ref{Fig_Characterization}c. We see good qualitative agreement with the DFT spectrum, although we have artificially increased the SOC strength in the tight-binding model for clarity.

We first characterized our samples by measuring the four-terminal electrical resistivity $\rho$ as a function of temperature $T$ with the current applied along the $c$-axis. Consistent with previous work \cite{Gooth2019}, $\rho$ increases with decreasing $T$ and its logarithmic derivative shows a peak around the expected CDW transition temperature of $T_\text{CDW} \sim$ 260 K (Fig.\ref{Fig_Characterization}d). Moreover, the logarithmic derivative saturates around a value of 0.7 (10\textsuperscript{-3} K), corresponding to a gap size of 250  meV. This is also consistent with previous transport experiments \cite{Gooth2019}. 

\begin{figure*}[ht!]
\includegraphics[width=2\columnwidth]{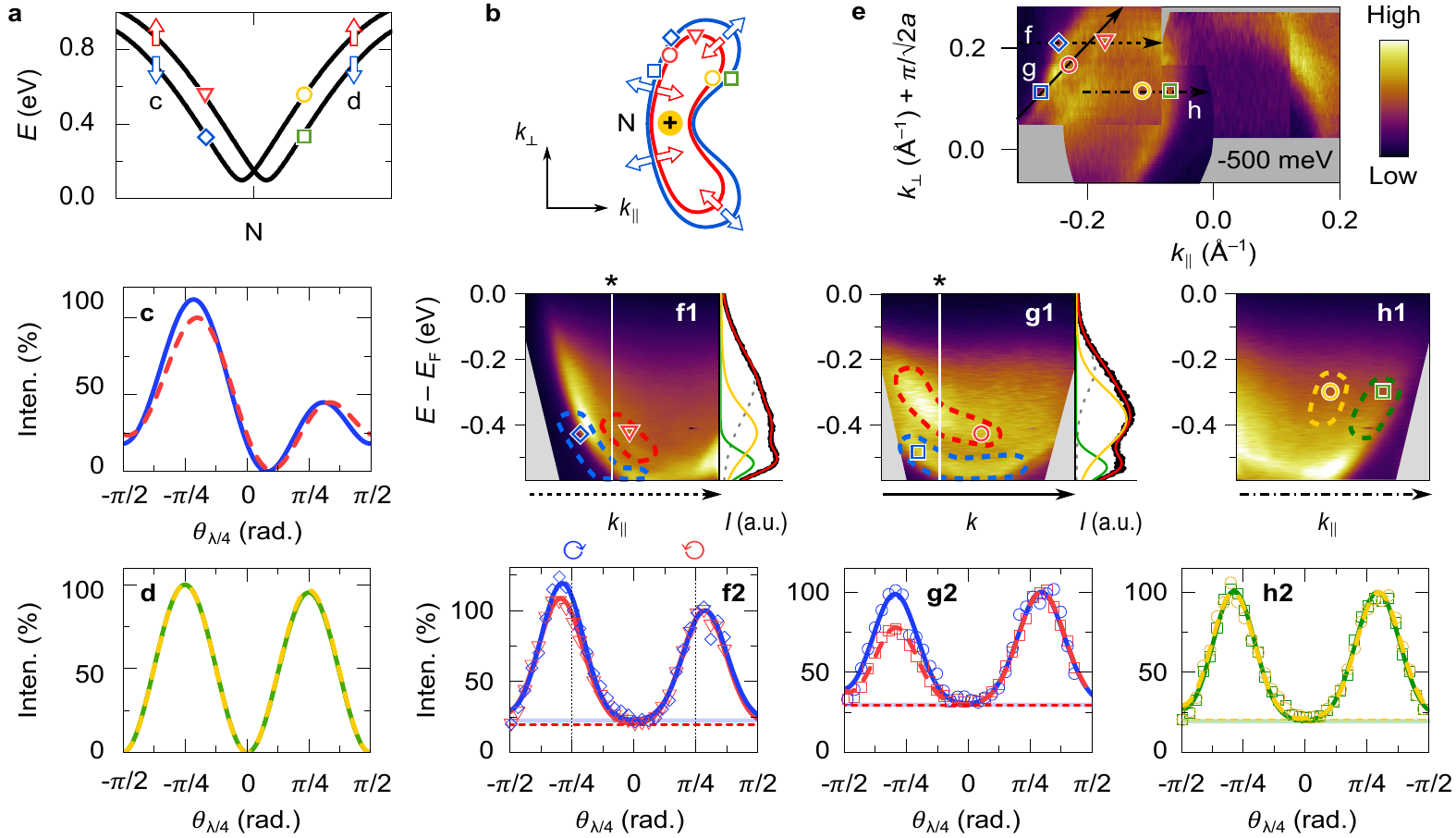}
\caption{\textbf{Light helicity dependent ARPES around the N point in (TaSe$_4$)$_2$I.} \textbf{(a)} Calculated tight-binding (TB) model band structure near the N point along the N'-N-N' path. Each Kramers-pair band is separated by spin-orbit coupling (SOC) except at the TRIM point. Arrows indicate up/down (pseudo-)spin. \textbf{(b)} Sketch of a constant energy map and the (pseudo-)spin texture around the N Kramers-Weyl point. \textbf{(c,d)} Calculated $\lambda$/4 waveplate angle ($\theta_{\lambda/4}$) dependence of photoemission intensities in the TB model for bands to the left (c) and right (d) of the N point as shown in (a). (pseudo-)spin-up (down) bands that are located at higher (lower) energy are shown as dashed (solid) lines. \textbf{(e)} ARPES constant energy cut measured around the N point at $E-E_F = -500$ meV. \textbf{(f1-h1)} ARPES spectra ($E-E_F$ vs. $k$) along the arrows marked in (d). Energy distribution curves (at momentum marked with an asterisk) fitted with Voigt functions are displayed on the right side of f1 \& g1. \textbf{(f2-h2)} Integrated photoemission intensity as a function of $\theta_{\lambda/4}$ for each spin split band around the N point. Region of integration is shown by the dashed lines in (f1-h1). The photoemitting beam helicity is left or right-handed circular polarization at $\theta_{\lambda/4}=+\pi/4$ and $-\pi/4$ radians, respectively. Experimental data (symbols) are normalized to maximum intensity for $\theta_{\lambda/4}>0$. Solid lines are guides to the eye.}\label{Fig_qwdep}
\end{figure*}

To further characterize the CDW order, we also performed X-ray diffraction (XRD) as a function of temperature above and below $T_\text{CDW}$. As can be seen in Fig.~\ref{Fig_Characterization}e, satellite peaks corresponding to $q_\text{CDW}$ emerge for $T<T_\text{CDW}$, similar to observations from other scattering studies \cite{FavreNicolin2001, Fujishita1986}. The intensities of the CDW peaks follow the expected mean-field behavior with $T_\text{CDW} \sim$ 260 K (Fig.~\ref{Fig_Characterization}f). We also performed synchrotron-based ARPES experiments with a photon energy of 50 eV on the same set of samples to compare with previous such experiments. Figure~\ref{Fig_Characterization}g shows the in-plane band dispersion at room temperature along the chain direction (in this case the $\Gamma Z$-direction). We observe the characteristic linearly dispersing valence bands with a minimum at $\Gamma$ as theoretically predicted and seen by various ARPES experiments\cite{TournierColletta2013, TSItype3,Yi2021,Wujun2021}. Note that the valence band maximum is significantly below the chemical potential, indicating the intrinsic n-doping of (TaSe$_4$)$_2$I samples. At these high photon energies, the photoemission intensity for the conduction bands is much weaker compared to that of the valence bands\cite{Wujun2021}. To study the conduction band in more detail we performed laser ARPES experiments.

Our laser ARPES experimental configuration is shown in Fig.~\ref{Fig_TSI_bigmap}a along with the bulk 3D BZ. For a photon energy of 6 eV, photoemission primarily originates from the plane corresponding to $k_n \sim \text{0.6 } (2\pi/\sqrt2a)$ \cite{Suppl}. This plane is close to the high symmetry N points, as indicated by the shaded plane in Fig.~\ref{Fig_TSI_bigmap}a. For our measurements, the sample is aligned such that the Ta chains are parallel to the analyzer slit which gives energy as a function of $k_\parallel=k_z$ for a given scan. The sample is then rotated so that $k_\perp$ is close to the momentum of the N TRIM point. Changes in $k_\perp$ around this point are measured by using electronic deflection without sample rotation. 
We first show constant energy ARPES maps corresponding to energies $E - E_F$ = $-550$ meV and $-500$ meV in Figs.~\ref{Fig_TSI_bigmap}b and \ref{Fig_TSI_bigmap}c, respectively. As shown in Fig.~\ref{Fig_TSI_bigmap}b, two bands are observed around the Kramers-Weyl node at the N TRIM point, i.e., around the coordinate (0.5, 0.5) in units of ($2\pi/c$, $2\pi/\sqrt{2}a$). These bands disperse outward with increasing kinetic energy, i.e.\ decreasing binding energy, indicating that they correspond to the conduction band of (TaSe$_4$)$_2$I near the Kramers-Weyl node at the N TRIM point when compared with first-principles calculations and our tight-binding model (Fig.~\ref{Fig_Characterization}b,c). The almost linearly dispersing `V-shaped' conduction bands are seen more clearly in the energy vs.\ momentum ($k_\parallel$) cut (see Fig.~\ref{Fig_TSI_bigmap}d for a plot along the white lines illustrated in \ref{Fig_TSI_bigmap}c). Similar `V-shaped' bands with relatively high velocities were also resolved in Refs.~\cite{Wujun2021} and \cite{Yi2021} for momentum along $k_\parallel$. On the other hand, for momentum along $k_{\perp}$ (Figs.~\ref{Fig_TSI_bigmap}e and \ref{Fig_TSI_bigmap}f), the bands have a relatively flat dispersion. These weakly dispersing bands along the $k_{\perp}$ direction are a characteristic feature of (TaSe$_4$)$_2$I due to its one-dimensional nature and have been observed in a number of previous ARPES studies \cite{TournierColletta2013, TSItype3,Yi2021}. We note that in our ARPES data, the chemical potential ($\mu$) of (TaSe$_4$)$_2$I is lower than that of a reference sample (Au or Bi$_2$Se$_3$). The top of the occupied bands is about 100 meV below $\mu$. As established in early ARPES works \cite{Dardel1991, Perfetti2001}, this is due to a strong polaronic effect which makes the spectral weight near the chemical potential incoherent. 

Having located the conduction bands originating from a predicted Kramers-Weyl node in (TaSe$_4$)$_2$I, we now characterize their spin texture using helicity-dependent laser ARPES measurements. In general, Kramers-Weyl nodes can be distinguished from conventional band-inversion Weyl nodes by their spin texture \cite{Hasan2018}. Construction of any Fermi surface enclosing a single Kramers-Weyl fermion maps onto itself under time-reversal symmetry. Since time-reversal also flips spin, this constrains the electronic states on opposite sides of the Fermi surface around a Kramers-Weyl node to have opposite spin. Note that this is in contrast to a conventional Weyl semimetal arising from band inversion, where there are no symmetry constraints on states at a single Fermi surface. The presence of additional rotational symmetries can further constrain the spins of states near the Kramers-Weyl node; in the isotropic limit, we expect to see an approximately radial spin texture arising from the dominant $\textit{\textbf{k}}\cdot\boldsymbol{\sigma}$ term in the Kramers-Weyl Hamiltonian. This was recently observed in spin-resolved ARPES experiments near the Kramers-Weyl points in elemental tellurium \cite{Gatti2020,Sakano2020}. A similar approximately radial spin texture is expected around Kramers-Weyl nodes of (TaSe$_4$)$_2$I, albeit with a stronger anisotropy due to its quasi-one-dimensional band structure\cite{Suppl}.

Due to the nature of the spin texture around the observed N KW point, we expect a distinctive asymmetry in the helicity-dependent photoemission from bands on either side of this point. This is illustrated in Fig.~\ref{Fig_qwdep}a-d (full details on the calculation in the SI \cite{Suppl}). Figures~\ref{Fig_qwdep}c and \ref{Fig_qwdep}d show the calculated photoemission intensity as a function of light-helicity on the left and right side of the N point respectively, as shown in Fig.~\ref{Fig_qwdep}a. Note that there is a clear difference in the photoemission intensities for opposite helicities of light on one side of the N point but not on the other. We attribute this to a combination of the chirality of the crystal and the incidence angle of the applied light. Since the crystal is chiral, electronic states on the two sides of the N point are related by a twofold rotation symmetry and by time-reversal symmetry. Both of these symmetry operations change the polarization and incidence angle of the incoming light, leading to an asymmetry in the photoemission matrix elements.  

\begin{figure}[ht!]
\includegraphics[width=0.85\columnwidth]{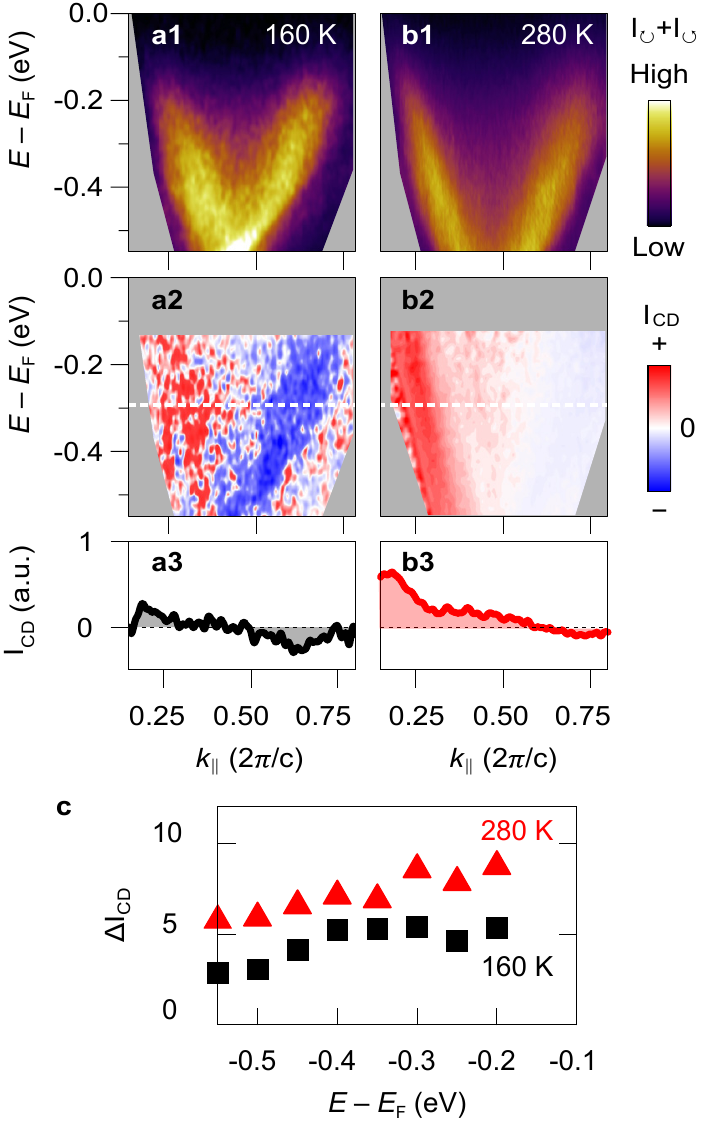}
\caption{\textbf{CD-ARPES spectra of (TaSe$_4$)$_2$I across the CDW phase transition near the N point.}ARPES scan near $k_{\parallel} = -0.5$ $ (2\pi/\sqrt{2}a)$ are taken with LHC ($I_{\circlearrowleft}$) and RHC ($I_{\circlearrowright}$) probe polarizations at (a) 160 K, and (b) 280 K. \textbf{(a1,b1)} The sum of both polarizations, $\text{I}_{\circlearrowright}+\text{I}_{\circlearrowleft}$. \textbf{(a2,b2)} Circular dichroism intensity maps I$_\text{CD}$ =  $(\text{I}_{\circlearrowright}-\text{I}_{\circlearrowleft})/(\text{I}_{\circlearrowright}+\text{I}_{\circlearrowleft})$. \textbf{(a3,b3)} A cut of normalized dichroism intensity, I$_\text{CD}$ =  $(\text{I}_{\circlearrowright}-\text{I}_{\circlearrowleft})/(\text{I}_{\circlearrowright}+\text{I}_{\circlearrowleft})$, taken at $E-E_F$ = -0.3 eV (white dashed line in (a2,b2)), and summed over a width of 20 meV. \textbf{(c)} Temperature-dependent $\Delta$I$_\text{CD}$, the intensity difference between minimum and maximum I$_\text{CD}$.}\label{Fig_CD_Tdep}
\end{figure}

To study if this is indeed the case, we modulated the light-helicity of the photoemitting beam using a quarter waveplate (angle denoted by $\theta_{\lambda/4}$). We isolate the bands around the N point (Fig.~\ref{Fig_qwdep}e) and plot the integrated spectral intensity in each band (region of integration is shown by the dashed contours in Fig.~\ref{Fig_qwdep}f1-h1) as a function of $\theta_{\lambda/4}$. The results are shown in Fig.~\ref{Fig_qwdep}f2-h2. The photoemission helicity dependence on one side of the KW point (Fig.~\ref{Fig_qwdep}f2,g2) is significantly more asymmetric than the other side (Fig.~\ref{Fig_qwdep}h2) in agreement with theoretical predictions (Fig.~\ref{Fig_qwdep}c,d). In addition, each of the spin-split bands on one side of the N point has the same helicity dependence whereas that is not the case for the opposite side. There is a clear difference in the observed intensity between left and right circularly polarized light as shown in Fig.~\ref{Fig_qwdep}f2,g2 (The spin-split bands can clearly be identified in the EDC cuts in Fig.~3f1,g1). These observations are a direct consequence of the presence of KW fermions in (TaSe$_4$)$_2$I and might also explain the observed circular dichroism in other photoemission experiments on (TaSe$_4$)$_2$I using higher photon energies \cite{Yi2021}.

We note here that the $\theta_{\lambda/4}$-dependence of the photoemission intensity near a KW point of (TaSe$_4$)$_2$I is uniquely related to the radial (pseudo-)spin texture around the KW point and is quite different from other well-studied systems such as topological insulators and strong Rashba SOC materials with a tangential spin texture \cite{gedik2011, Fu2009, Jung2011, Ryu2017,Crepaldi2014}. For those systems, circular dichroism (CD-ARPES) experiments are typically performed to measure the photoemission intensity difference between left and right circularly polarized light which in turn gives a measure of the pseudospin texture \cite{Chiang2011}. To compare our helicity dependent measurements on (TaSe$_4$)$_2$I with those on a system with a tangential pseudo spin texture, we studied the prototypical topological insulator Bi$_2$Se$_3$ (see Fig.~S2 in \cite{Suppl} for more details) with our setup. As observed in previous measurements\cite{Chiang2011, gedik2011, Fu2009, Jung2011}, we obtain a symmetric $\theta_{\lambda/4}$ dependence of the photoemission intensity at points on either side of the Dirac point. This is in contrast to the asymmetric intensity observed in (TaSe$_4$)$_2$I for bands near the N point. 
 
We next perform circular dichroism (CD) ARPES ($I_{\circlearrowleft}$ - $I_{\circlearrowright}$) both above and below $T_\text{CDW}$ near the N KW point. In general, the CDW order nests and gaps out the FSWPs in (TaSe$_4$)$_2$I as detailed in Ref.\cite{Wujun2021}. This gapped Weyl-CDW phase is realized by connecting FSWPs with opposite chiral charges \cite{Bobrow2020}. Note that there are 24 pairs of FSWPs in a bulk BZ that are quite close to each other in both energy and momentum space \cite{Wujun2021}, which makes it difficult to resolve individual FSWPs even with state-of-the-art ARPES setups \cite{Sobota2021, Zhou2018}. In contrast, KW points with opposite chirality are well separated in both energy and momentum \cite{Hasan2018}. Whether the onset of CDW order in (TaSe$_4$)$_2$I affects the spin texture around the KW points is an open question. 

Figure~\ref{Fig_CD_Tdep} presents CD-ARPES in (TaSe$_4$)$_2$I both above and below $T_\text{CDW}$ around the N point. Note that for $T < T_\text{CDW}$, the linearly dispersing bands in Fig.~\ref{Fig_CD_Tdep}a1 meet at a slightly higher energy when compared with the spectra for $T > T_\text{CDW}$. This indicates the opening up of a gap around $E-E_F = -0.6$ eV, consistent with previous ARPES studies \cite{TournierColletta2013, Wujun2021}. Figures~\ref{Fig_CD_Tdep}a2 and \ref{Fig_CD_Tdep}b2 show the resulting CD-ARPES spectra. The bands on the opposite sides of the N point have opposite spins as expected from our helicity dependent studies above. This is also highlighted in Fig.~\ref{Fig_CD_Tdep}a3,b3 where the sign of $I_\text{CD}$ flips across the N point. Furthermore, our measurements reveal a decrease in the overall circular dichroism below $T_\text{CDW}$, as shown in Fig.~\ref{Fig_CD_Tdep}c, suggesting that CDW order mixes the chirality of the KW points in addition to mixing the chirality of FSWPs. Certainly, significant band renormalization (and thus mixing of Weyl point chirality) is expected in the vicinity of the CDW gap though it is unclear whether this mixing should prevail up to the higher energies of the KW fermions near the N point. 

In conclusion, by using helicity-dependent and CD-APRES, ab-initio calculations, and tight-binding modeling, we have investigated the KW fermions at the N point of (TaSe$_4$)$_2$I both above and below the CDW ordering transition. Our work provides the first experimental evidence for the presence of KW fermions in this material. Furthermore, the change in strength of the CD-ARPES signal across the CDW ordering temperature points towards the impact of the phase transition on quasiparticle band topology. Our results suggest that a deeper theoretical and experimental investigation of helicity-dependent and CD-ARPES in the ordered phase could shed light on the exotic properties predicted for this Weyl-CDW compound.

\vspace{2mm}
\noindent\textbf{METHODS}

\noindent\textit{Sample preparation.} 
Single crystals were prepared by adapting a chemical vapor transport technique reported by Maki \textit{et al.}\cite{maki_charge_1983} Stoichiometric amounts of Ta wire (99.9\%), Se powder (99.999\%) and I shot (99.99\%) were loaded into  a fused silica tube, which was sealed under vacuum and heated with a source temperature of 600$^\circ$C and sink temperature of 500$^\circ$C for 10 days. 
X-ray diffraction patterns were collected on a Bruker D8 ADVANCE diffractometer with Mo $K\alpha$ radiation. Resistivity was measured in four-point geometry in a Quantum Design Physical Property Measurement System.

\noindent\textit{Density-functional theory.}
We perform fully relativistic, non-collinear first-principles density functional theory (DFT) \cite{Hohenberg:1964} simulations, including the spin-orbit interaction, using the Vienna \emph{Ab-Initio} Simulation Package (VASP) \cite{Kresse:1996,Kresse:1999,Gajdos:2006,Steiner:2016}.
We converted the atomic coordinates of the conventional unit cell of (TaSe$_4$)$_2$I provided by Materials Project (mpID 30531) \cite{Jain2013,Ong2012b} to a primitive unit cell using AFLOW/ACONVASP \cite{Setyawan:2010}.
The generalized-gradient approximation (GGA) as parameterized by Perdew, Burke, and Ernzerhof (PBE)  \cite{Perdew:1997} was used to describe exchange and correlation.
Kohn-Sham states were expanded into a plane-wave basis with a kinetic-energy cutoff of 520 eV.
A $10\times 10\times 10$ $\Gamma$-centered Monkhorst-Pack grid \cite{Monkhorst:1976} was used for Brillouin zone sampling and the resulting Kohn-Sham Hamiltonian was diagonalized for on finely sampled high-symmetry lines in reciprocal space to obtain the electronic structure data in this work.

\noindent\textit{Single crystal X-ray scattering.} Single crystal X-ray scattering measurements were carried out using the in-lab X-ray instrument equipped with a Xenocs GeniX3D Mo K$\alpha$ microspot X-ray source with multilayer focusing optics, providing $2.5 \times 10^7$ photons/sec in a beam spot of 130 $\mu$m at the sample position. The samples were cooled by a closed-cycle helium cryostat with a base temperature of 8 K mounted to a Huber four-circle diffractometer. The momentum resolution varied between $\Delta q$ = 0.01 \AA$^{-1}$ and 0.08 \AA$^{-1}$ depending on the location in momentum space. Scattering signals were collected by a Mar345 image plate detector with 3450×3450 pixels. Three-dimensional surveys of momentum space were performed by taking images in 0.05$^{\circ}$ increments while sweeping samples through an angular range of 20$^{\circ}$ and mapping each pixel to the corresponding location in momentum space.

\noindent\textit{Synchrotron ARPES.} Measurements were performed using $p$-polarized 50-eV photons and a Scienta R4000 energy analyzer with a step size of 15 meV at Beamline 10.0.1.1, Advanced Light Source. The (TaSe$_4$)$_2$I crystal was cleaved \textit{in-situ} in an ultra high vacuum environment at room temperature. The analyzer slit was set parallel to the $\Gamma$Z direction.

\noindent\textit{Laser ARPES.} The light  helicity-dependent ARPES measurements were performed using a hemispherical analyzer with electronic deflection (Scienta Omicron, DA30-L) to map both $k_x$ and $k_y$ without sample rotation. The photon energy was set to 6 eV (206 nm) using a custom-built setup that generated the 5th harmonic of a 1030 nm beam from a Yb-based laser sources (Light Conversion PHAROS). The plane of incidence is set to the $k_\text{n}$-$k_\perp$ plane with a 45$^{\circ}$ incident angle (see Fig. S2 in \cite{Suppl}). The light helicity was adjusted with a quarter waveplate. The entrance slit of the hemispherical analyzer was parallel to the chain direction of the sample. 
The dichroism plots in Fig. \ref{Fig_CD_Tdep} are normalized with the sum, $I_\text{CD}$ =  $(I_{\circlearrowright}-I_{\circlearrowleft})/(I_{\circlearrowright}+I_{\circlearrowleft})$ to compensate for laser instability and aging effects over time.

\noindent\textbf{Acknowledgements.} We thank Benjamin Wieder for fruitful discussions. This study was supported by the Center for Quantum Sensing and Quantum Materials, an Energy Frontier Research Center funded by the U. S. Department of Energy, Office of Science, Basic Energy Sciences under Award DE-SC0021238.
This research used resources of the Advanced Light Source (ALS), which is a DOE Office of Science User Facility under Contract No.\ DE AC02 05CH11231.
NB and AS acknowledge support from the Illinois Materials Research Science and Engineering Center, supported by the National Science Foundation MRSEC program under NSF Award No.\ DMR-1720633.
This work made use of the Illinois Campus Cluster, a computing resource that is operated by the Illinois Campus Cluster Program (ICCP) in conjunction with the National Center for Supercomputing Applications (NCSA) and which is supported by funds from the University of Illinois at Urbana-Champaign.
The crystal structure plotted in Fig. \ref{Fig_Characterization}a was generated with the VESTA software \cite{Momma2011}.

\noindent\textbf{Data availability.} All relevant data are available on reasonable request.

\noindent\textbf{Competing financial interests:} The authors declare no competing financial interests.

\noindent\textbf{Author contributions:} S.K., N.B. and F.M. performed the laser-ARPES experiments and the corresponding data analysis. R.C.M and B.B. developed the theoretical methods, the tight-binding model and the theoretical analysis. A.S. developed and carried out the DFT simulations. C.Z. and D.P.S synthesized the samples and performed the four terminal resistivity measurements. M-K.L, J.A.H, S-K.M and T.-C.C. performed the synchrotron ARPES experiments. X.G. and P.A. carried out the XRD measurements. S.K., R.C.M, B.B. and F.M. wrote the manuscript with input from all the authors. This project was supervised and directed by B.B. and F.M.

\clearpage
\newpage

\renewcommand{\thefigure}{S\arabic{figure}}
\setcounter{figure}{0}    

\section{Supplementary Materials}

\subsection{I. A Tight-Binding Model for (TaSe$_4$)$_2$I}
In this section we present a tight-binding model for (TaSe$_4$)$_2$I. Though this model is overly simplistic and does not capture the full phenomenology of the real material, we believe it captures several generic features of \tsi in the weakly-coupled-chain approximation, and hence should allow us to make qualitatively correct statements about experiments near the $N$ point.

To begin, we recall that \tsi crystalizes in space group $I4221'$ (\# 97). Information about the symmetry generators, Wyckoff position, and $\mathbf{k}$-point labels for this space group can be obtained from the Bilbao Crystallographic Server\cite{aroyo2006bilbaoa,aroyo2006bilbao,aroyo2011crystallography}. The fourfold axis is, conventionally, taken to be the $c$-axis. We take as our basis of the Bravais lattice the vectors
\begin{align}
    \mathbf{e}_1&= \frac{1}{2}(-a\mathbf{\hat{x}}+a\mathbf{\hat{y}}+c\mathbf{\hat{z}}), \\
    \mathbf{e}_2&= \frac{1}{2}(a\mathbf{\hat{x}}-a\mathbf{\hat{y}}+c\mathbf{\hat{z}}), \\
    \mathbf{e}_3&= \frac{1}{2}(a\mathbf{\hat{x}}+a\mathbf{\hat{y}}-c\mathbf{\hat{z}}),
\end{align}
where, for the real material the lattice constants take the values $a\approx 9.5\AA$ and $c\approx 12.8\AA$. The reciprocal lattice vectors are, accordingly
\begin{align}
    \mathbf{g}_1&=2\pi\left(\frac{1}{a}\mathbf{\hat{y}}+\frac{1}{c}\mathbf{\hat{z}}\right), \\
    \mathbf{g}_2&=2\pi\left(\frac{1}{a}\mathbf{\hat{x}}+\frac{1}{c}\mathbf{\hat{z}}\right), \\
    \mathbf{g}_3&=\frac{2\pi}{a}\left(\mathbf{\hat{x}}+\mathbf{\hat{y}}\right).
\end{align}
According to the topological materials database\cite{2019topological,vergniory2021all}, this material is a compatibility-relation enforced semimetal, with band crossings along the line $\Lambda = (u,u,-u)$ (in reduced coordinates). In terms of little group representations, this is enforced by the fact that the occupied representations at $\Gamma$ and $M$ closest to the Fermi level are, respectively, $\bar{\Gamma}_7$ and $\bar{M}_6$, while the first unoccupied representations are $\bar{\Gamma}_6$ and $\bar{M}_7$. This, along with the chemical structure, is consistent with a low-energy model focusing on Ta atoms located  at the 4c Wyckoff position, with representative coordinates
\begin{align}
    \mathbf{r}_0 &= \frac{1}{2}(\mathbf{e}_1 + \mathbf{e}_3), \\
    \mathbf{r}_1 &=-\frac{1}{2}(\mathbf{e}_2+\mathbf{e}_3).
\end{align}

The site-symmetry group of the 4c position is the point group $222$, which has the nice feature that it has only one irreducible representation, labeled $\bar{E}$, which we may as well write in terms of spinful s-orbitals for convenience. We will thus construct a tight-binding model for spinful s-orbitals at the 4c position.

In order to construct our tight-binding Hamiltonian, we will need to know the action of the space group symmetry operations on the positions of the electronic orbitals. We make use of the following relations:
\begin{align}
    C_{4z}\mathbf{r}_0 &= \mathbf{r}_1,\;\; C_{4z} \mathbf{r}_1 = \mathbf{r}_0-\mathbf{e}_1-\mathbf{e}_3, \\
    C_{2x}\mathbf{r}_0 & = \mathbf{r}_0 - \mathbf{e}_1-\mathbf{e}_3, \;\; C_{2z}\mathbf{r}_1 = \mathbf{r}_1.
\end{align}

To make contact with the quasi-one-dimensional structure of the real material, we should expect hopping along the $\mathbf{z}$-axis (the chain direction) to dominate over other hopping terms. Furthermore, note that because of the fourfold rotational symmetry, any spin-dependent hopping along the $z$-direction must, in this model, be proportional to $\sigma_z$. Note, also, that the shortest $z$-directed hopping is from $\mathbf{r}_0$ to $\mathbf{r}_1+\mathbf{e}_1+\mathbf{e}_2+\mathbf{e}_3$, along with its orbits under the point group symmetries. We can thus write the spin-independent and spin dependent $z$-directed hoppings
\begin{align}
    H_1 = t_1\sum_\mathbf{R} c^\dag_{0\mathbf{R}}c_{1\mathbf{R+e_1+e_2+e_3}} + c^\dag_{0\mathbf{R}}c_{1\mathbf{R+e_3}}+\mathrm{h.c.}\label{eq:h1}
\end{align}
and
\begin{align}
    H_{\lambda_1} = i\lambda_1\sum_\mathbf{R} c^\dag_{0\mathbf{R}}\sigma_zc_{1\mathbf{R+e_1+e_2+e_3}} - c^\dag_{0\mathbf{R}}\sigma_zc_{1\mathbf{R+e_3}}+\mathrm{h.c.}\label{eq:hld1}
\end{align}
where $c_{i\mathbf{R}}$ is the annihilation operator for an electron at site $\mathbf{r}_i$ in unit cell $\mathbf{R}$, and we have suppressed the spin indices for brevity. Using the Fourier transform convention
\begin{equation}
c_{i\mathbf{R}} = \sum_\mathbf{k} e^{i\mathbf{k}\cdot(\mathbf{R}+\mathbf{r}_i)}c_{i\mathbf{k}},
\end{equation}
we can rewrite Eqs.~(\ref{eq:h1}) and (\ref{eq:hld1}) as
\begin{align}
    H_1+H_{\lambda_1} &=2t_1\sum_{ij\mathbf{k}}  c^\dag_{i\mathbf{k}} \cos\frac{k_1+k_2}{2}\tau_x^{ij} c_{j\mathbf{k}}\nonumber \\
    &+2\lambda_1\sum_{ij\mathbf{k}}  c^\dag_{i\mathbf{k}}\sigma_z\sin\frac{k_1+k_2}{2}\tau_x^{ij} c_{j\mathbf{k}}
\label{eq:zham}
\end{align}
where we have introduced a set of Pauli matrices $\vec{\tau}$ which act in the space of orbitals $(\mathbf{r}_0,\mathbf{r}_1$), and we have defined $k_i = \mathbf{k}\cdot\mathbf{e}_i$. We see directly from Eq.~(\ref{eq:zham}) that $H_1+H_{\lambda_1}$ consists entirely of hopping along the $k_z=k_1+k_2$ direction, and so is the lowest-order Hamiltonian describing the decoupled (TaSe$_4$) chains in (TaSe$_4$)$_2$I.  
Furthermore, note that there is a twofold-degenerate nodal plane when $k_1+k_2 = \pm2\arctan (t_1/\lambda_1)$. The states in the nodal plane are eigenstates of $\sigma_z$: spin down eigenstates for the $+$ sign, and spin up eigenstates for the $-$ sign. 

We will now add interchain couplings to our model, which will provide a nontrivial dispersion for Kramers-Weyl Fermions at the TRIM points. Rather than providing an exhaustive catalogue of couplings, we will limit ourselves to adding only enough terms to the Hamiltonian to get a linear Weyl dispersion at the $N$ point, which we will verify below.  We first consider spin-independent hoppings from $\mathbf{r}_i$ to $\mathbf{r_i+e_1+e_2}$, and from $\mathbf{r}_i$ to $\mathbf{r_i+e_2+e_3}$. We write
\begin{equation}
    H_3 = t_3\sum_{\mathbf{Ri}}(-1)^i(c^\dag_{i\mathbf{R}}c_{i\mathbf{R+e_1+e_3}}-c^\dag_{i\mathbf{R}}c_{i\mathbf{R+e_2+e_3}})+\mathrm{h.c.}
\end{equation}
which yields the Bloch Hamiltonian
\begin{equation}
    H_3=2t_3\sum_{ij\mathbf{k}}(\cos k_1+k_3 - \cos k_2+k_3)c^\dag_{i\mathbf{k}}\tau_z^{ij}c_{j\mathbf{k}}.
\end{equation}

\begin{widetext}
Finally, we consider inter-chain spin-orbit coupling. We find the three simplest such terms are
\begin{align}
    H_{\lambda_2}&=i\lambda_2\sum_\mathbf{R}c^\dag_{0\mathbf{R}}\left[(\sigma_x+\sigma_y)c_{1\mathbf{R+e_1+e_2+2e_3}}
    +(\sigma_x-\sigma_y)c_{1\mathbf{R+e_2+e_3}}
    +(\sigma_y-\sigma_x)c_{1\mathbf{R+e_1+e_3}}
    -(\sigma_x+\sigma_y)c_{1\mathbf{R}}\right]+\text{h.c.},\\
    H_{\lambda_3}&=i\lambda_3\sum_\mathbf{R}c^\dag_{0\mathbf{R}}\sigma_z\left[ c_{1\mathbf{R+e_1+e_2+2e_3}}
    -c_{1\mathbf{R+e_2+e_3}}
    -c_{1\mathbf{R+e_1+e_3}}
    +c_{1\mathbf{R}}\right]+\text{h.c.},
    \\
    H_{\lambda_4}&=i\lambda_4\sum_\mathbf{R}c^\dag_{0\mathbf{R}}\sigma_yc_{0\mathbf{R+e_1+e_3}} + c^\dag_{1\mathbf{R}}\sigma_xc_{1\mathbf{R+e2+e3}} + \text{h.c.} ,
\end{align}
which can be written in momentum space as
\begin{align}
    H_{\lambda_2}(\mathbf{k})&=-2\lambda_2\sum_\mathbf{k}c^\dag_{i\mathbf{k}}\left[\sigma_x\left(\sin\frac{k_1+k_2+2k_3}{2} - \sin\frac{k_1\textcolor{blue}{-}k_2}{2}\right) +\sigma_y\left(\sin\frac{k_1+k_2+2k_3}{2} + \sin\frac{k_1\textcolor{blue}{-}k_2}{2}\right)\right]c_{j\mathbf{k}}\tau_x^{ij}, \\
    H_{\lambda_3}(\mathbf{k})&=4\sum_{\mathbf{k}}c^\dag_{i\mathbf{k}}\sigma_z\tau_y^{ij}\sin\frac{k_1+k_3}{2}\sin\frac{k_2+k_3}{2}c_{j\mathbf{k}},\\
    H_{\lambda_4}(\mathbf{k})&=2\lambda_4\sum_\mathbf{k}c^\dag_{0\mathbf{k}}\sigma_y\sin(k_1+k_3)c_{0\mathbf{k}} + c^\dag_{1\mathbf{k}}\sigma_x\sin(k_2+k_3)c_{1\mathbf{k}}.
\end{align}
\end{widetext}

\subsection{II. Kramers-Weyl Fermion at the $N$ Point}
Let us now turn to the $N$ point, with coordinates $N=\frac{1}{2}\mathbf{g}_2$. The little group of the $N$ point is generated by $C_{2y}$ and time-reversal symmetry. There is only one unique irreducible (co)representation at the $N$ point, with representation matrices
\begin{align}
    \rho(C_{2y}) &= i\mu_y, \nonumber \\
    \rho(TR)&=i\mu_y\mathcal{K},
\end{align}
where $\vec{\mu}$ is a vector of Pauli matrices in the pseudospin space. Due to the relative lack of symmetry constraints, the most general Hamiltonian for the Kramers Weyl degeneracy and $N$ takes the form
\begin{equation}
    H_N(\delta\mathbf{k}) = a\delta k_y\mu_y + (\delta k_x , \delta k_z)\left(\begin{array}{cc} b & c \\
    d & e \end{array}\right)\left(\begin{array}{c} \mu_x \\ \mu_z\end{array}\right),
\end{equation}
where $a,b,c,d,e$ are model dependent constants. $H_N$ describes a Weyl node whenever $a\neq 0$ and $be-cd\neq 0$, which will generically be the case absent fine tuning.

Let us now turn to the Kramers Weyl fermions at the $N$ point in our tight-binding model. Taking again
\begin{equation}
    H=H_1+H_3+H_{\lambda_1}+H_{\lambda_2}+H_{\lambda_3}+H_{\lambda_4},
\end{equation}
we have that
\begin{equation}
H_N = 4t_3\tau_z + 2\lambda_1 \tau_x\sigma_z -4\lambda_2\tau_x\sigma_x.
\end{equation}
Let us rewrite this as
\begin{equation}
    H_N=4t_3\tau_z + 4\lambda\tau_x\hat{n}\cdot\vec{\sigma},
\end{equation}
where we have introduced
\begin{equation}
    \lambda=\sqrt{\lambda_2^2+\lambda_1^2/4}
\end{equation}
and
\begin{equation}
    \hat{n} = \frac{1}{\lambda}(-\lambda_2,0,\lambda_1/2).
\end{equation}
We see from this that the Kramers Weyl fermions have energies 
\begin{equation}
    E_\pm = \pm 4\sqrt{t_3^2+\lambda^2}.
\end{equation}
Focusing on the $+$ state, we see that it is spanned by the states
\begin{align}
    |\psi_1\rangle &= |\hat{m}\cdot\tau = +1, \hat{n}\cdot\sigma = +1\rangle, \nonumber \\
    |\psi_2\rangle &= |C_{2z}\hat{m}\cdot\tau = +1, \hat{n}\cdot\sigma = -1\rangle,
\end{align}
where we have introduced 
\begin{align}
    \hat{m} &= \frac{1}{\sqrt{t_3^2+\lambda^2}}(\lambda, 0, t_3), \\
        C_{2z}\hat{m} &= \frac{1}{\sqrt{t_3^2+\lambda^2}}(-\lambda, 0, t_3).
\end{align}
We see that at the N point there is nontrivial entanglement between spin and orbital degrees of freedom, such that we can no longer identify the $\mu$ degree of freedom purely with the electron spin. Nevertheless, we can linearize $H$ about the $N$ point, and focus on the line $k_x=k_y=0$ that is relevant for our experiment. We find that
\begin{align}
    H_N(k_z)\approx &\frac{2\pi\lambda_1t_1k_z}{E_+}(|\psi_1\rangle\langle\psi_1|- |\psi_2\rangle\langle\psi_2|) \nonumber \\
    &+ \frac{4\pi\lambda_2 t_1k_z}{E_+}(|\psi_1\rangle\langle\psi_2| + |\psi_2\rangle\langle\psi_1|).
\end{align}
Combining this with our expression for the eigenstates $|\psi_{1,2}\rangle$ in terms of the orbital and spin degree of freedom, we can derive an expression for the average spin $\langle \vec{\sigma}/2\rangle/ (k_z)$ for the conduction band near the Weyl point. We find
\begin{equation}
    \langle \vec{\sigma}/2\rangle(k_z) = \frac{\mathrm{sign}(k_z)}{2\sqrt{\lambda_1^2+4\lambda_2^2}}(-2\lambda_2,0,\lambda_1)=\frac{1}{2}\mathrm{sign}(k_z)\hat{n}.
\end{equation}
In the physical limit that $\lambda_1 \gg \lambda_2$, we see that we still expect to have spin polarization largely parallel to momentum.

\subsection{III. Dipole Matrix Element Derivation}
In this section, we compute the CD-ARPES matrix elements from our tight-binding model.  We begin with Fermi's Golden Rule for a transition between final and initial states with an interaction Hamiltonian, $V_I$ \cite{Cao2013, Xiao2010, xidaiDirac, weylxidai},
\begin{equation}
\label{FirstPrincipleTransitionRate}
    w_{i \rightarrow f}(\mathbf{k}) = \frac{2 \pi}{\hbar} |\langle \psi_f(\mathbf{k}) | V_{I} | \psi_i(\mathbf{k}) \rangle|^2.
\end{equation}
This transition rate is proportional to the photoemission intensity  $I = \frac{\hbar \omega}{\sigma} w_{i \rightarrow f}$, where $\omega$ is the incident light frequency, and $\sigma$ is the spot size.  As in the main text, we will normalize the final intensity by its maximum, so proportionality statements are sufficient.

In a traditional ARPES setup, a light source bombards a sample material with vector potential, $\mathbf{A}$, and interaction Hamiltonian $V_I \propto \mathbf{A} \cdot \mathbf{v}$, where $\mathbf{v}=\dot{\mathbf{r}}$ is the velocity operator.  This is obtained by expanding the kinetic energy in $\mathbf{A}$ with a minimal coupling $\mathbf{p} \rightarrow \mathbf{p} + \mathbf{A}$ and dropping higher orders in $\mathbf{A}$ \cite{Niu1996, Giraud2012, weylxidai}.  The input vector potential is taken to be  elliptically polarized, parametrized by a quarter waveplate angle, $\theta_{\text{qw}}$, a polar angle $\theta$, and an azimuthal angle $\phi$ consistent with the experimental setup described in the main text. The angle $\theta_{\text{qw}}$ controls the direction of the fast axis in our ellipse.  To describe the elliptical polarization in our coordinate system, first consider circular polarization, $\mathbf{A}_{\eta}$, where the relative phase angle is given through complex representation: $\hat{x} \rightarrow \hat{x}$, $\hat{y} \rightarrow i \hat{y}$.  Then the circularly polarized vector potential is \cite{xidaiDirac, weylxidai, Hwang2015}
\begin{equation}
    \mathbf{A}_{\eta} = A_0 \begin{bmatrix}
\cos(\theta) \cos(\phi) + i \eta \sin(\phi)  \\
\cos(\theta) \sin(\phi) - i \eta \cos(\phi) \\
-\sin(\theta)
\end{bmatrix}.
\end{equation}
The right/left hand polarization is respectively denoted by $\eta = \pm 1$.  The elliptical polarization is formed through
\begin{equation}
\resizebox{1\hsize}{!}{$
\begin{aligned}
    \mathbf{A} =& \frac{(1-i)}{2} \left( \cos^2(\theta_{\text{qw}}) + (1 + i)\cos(\theta_{\text{qw}}) \sin(\theta_{\text{qw}}) + i \sin^2(\theta_{\text{qw}}) \right) A_{+1}
    \\&
    + \frac{(1-i)}{2} \left( \cos^2(\theta_{\text{qw}}) - (1 + i)\cos(\theta_{\text{qw}}) \sin(\theta_{\text{qw}}) + i \sin^2(\theta_{\text{qw}}) \right) A_{-1}.
\end{aligned}
$}
\end{equation}

In the photoemission experiment, incoming ellipically polarized light interacts with the material via $V_I$, and ejects an electron from the material \cite{Schler2020, Schattke2003, Hasan2018}.  Consequently, the final state may be approximated as a free electron \cite{Cao2013, Han2012, Moser2017} with wavefunction $\langle \mathbf{r}| \psi_{f, s} (\mathbf{k}) \rangle \approx e^{i \mathbf{k} \cdot \mathbf{r}} | s \rangle$, where $s$ denotes spin up or down in the $\hat{z}$ basis.  However, the initial state of the electron in the sample should be treated more sensitively.  We start by expanding the Bloch wavefunction in the orbital-spin basis of our tight-binding model \cite{Giraud2012, Han2012}:
\begin{equation}
     u_{\mathbf{k}, \alpha}(\mathbf{r}) = 
    \frac{1}{\sqrt{N}}\sum_{\sigma, \mathbf{R}, s} u^{\sigma, s}_{\alpha \mathbf{k}} \phi_{\sigma}(\mathbf{R} + \mathbf{r}_{\sigma} - \mathbf{r})e^{i \mathbf{k} \cdot (\mathbf{R} + \mathbf{r}_{\sigma} - \mathbf{r})} | s \rangle.\label{eq:initialwavefn}
\end{equation}
In this equation, $\phi_{\sigma} (\mathbf{R} + \mathbf{r}_{\sigma} - \mathbf{r})$ denotes the orbital wavefunction, centered about $\mathbf{R} + \mathbf{r}_{\sigma}$. We will take this type of orbital to be an s-wave for simplicity, so that $\phi_{\sigma} (\mathbf{r}_{\sigma} + \mathbf{r}) \propto e^{|\mathbf{r}_{\sigma} + \mathbf{r}| / a_0}$, where $a_0 \approx a / 5$ is the characteristic size of the orbital, with $a_0$ being the Bohr radius.  Ordinarily, there would be a spread in the wavefunction, namely, $\langle \mathbf{r} || \psi_{i, \alpha} \rangle = \int [d \mathbf{q}] w(\mathbf{k}, \mathbf{q}) e^{i \mathbf{q} \cdot \mathbf{r}} u_{\mathbf{q}, \alpha}(\mathbf{r})$ where $w(\mathbf{k}, \mathbf{q})$ is some spread function (e.g. a Gaussian) \cite{Niu1996, Schler2020, Xiao2010}.  However, we take this spread to be sufficiently peaked so that $w(\mathbf{k}, \mathbf{q})$ approximately follows a Dirac delta distribution.

We will now express the intensity in terms of $\mathbf{r}$.  Specifically, we can use the equations of motion $\mathbf{v} =\frac{-i }{\hbar} [\mathbf{r}, H]$ so that the (spin-resolved) intensity becomes $I_s \propto | (E_{f, s} - E_i)|^2|\langle \psi_{f, s}(\mathbf{k})| \mathbf{A} \cdot \mathbf{r} | \psi_i(\mathbf{k}) \rangle|^2$ \cite{Schattke2003, Hwang2015}.  For convenience, define the dipole matrix as $M_s \equiv \langle \psi_{f, s}(\mathbf{k})| \mathbf{A} \cdot \mathbf{r} | \psi_i(\mathbf{k}) \rangle$.  The outgoing momentum $\mathbf{p}$ is described by the conservation law $\mathbf{p} = \mathbf{k} + n_1 \mathbf{g}_1 + n_2 \mathbf{g}_2 + n_3 \mathbf{g}_3 + \mathbf{p}_{\bot}$, where $\mathbf{p}_{\bot}$ is the momentum correction due to overcoming the surface potential, and $\{n_1, n_2, n_3 \}\in \mathbb{Z}$ is the number of times the free-particle momentum gets backfolded into the first Brillioun zone \cite{Mahan1970}.  Since the emitted electron should be primarily perpendicular to the (110) crystallographic plane as shown in Fig. 1(a) in the main text, then $n_3 =0$ and $n_1 = -n_2 = n$.  This aligns $n(-\mathbf{g}_1 + \mathbf{g}_2)$ with the $\mathbf{k}_n$ direction.  Also, all the circular dichroic results are nearly the same for $|\mathbf{p}_{\bot}| \ll |\mathbf{k} + n (-\mathbf{g}_1 + \mathbf{g}_2)|$.  Therefore, the approximation $\mathbf{p} \approx \mathbf{k} + n (-\mathbf{g}_1 + \mathbf{g}_2)$ is valid \cite{Schler2020}. Combining this with the form of the initial state Eq.~(\ref{eq:initialwavefn}), we find
\begin{widetext}
\begin{equation}
\begin{split}
    M_s = & \langle \psi_{f, s}(\mathbf{k})| \mathbf{A} \cdot \mathbf{r} | \psi_i(\mathbf{k}) \rangle
    \\ = & 
     \sum_{\mathbf{R}, s^{\prime}, i}\int d\mathbf{r} \langle s | e^{-i (\mathbf{k} + n (-\mathbf{g}_1 + \mathbf{g}_2)) \cdot \mathbf{r}}\mathbf{A} \cdot \mathbf{r} u^{i, s^{\prime}}_{\alpha, \mathbf{k}} \phi_{\alpha}(\mathbf{R} + \mathbf{r}_{i} - \mathbf{r}) e^{i \mathbf{k} \cdot (\mathbf{R} + \mathbf{r}_i)}  | s^{\prime} \rangle
    \\ = & 
    \sum_{\mathbf{R},  i}\int d\mathbf{r}  e^{-i (\mathbf{k} + n (-\mathbf{g}_1 + \mathbf{g}_2)) \cdot (\mathbf{r} + \mathbf{R})} e^{i\mathbf{k} \cdot(\mathbf{R} + \mathbf{r}_i)} \mathbf{A} \cdot (\mathbf{r} + \mathbf{R}) \left(u^{i, s}_{\alpha, \mathbf{k}} \phi_{\alpha}(\mathbf{r} - \mathbf{r}_i)  \right)
    \\ = & 
   \sum_{\mathbf{R}, i}\int d\mathbf{r}  e^{-i \mathbf{k} \cdot (\mathbf{r} - \mathbf{r}_i)  -i n (-\mathbf{g}_1 + \mathbf{g}_2) \cdot (\mathbf{r} + \mathbf{R})  } \mathbf{A} \cdot (\mathbf{r} + \mathbf{R}) \left(u^{i, s}_{\alpha, \mathbf{k}} \phi_{\alpha}(\mathbf{r} - \mathbf{r}_i) \right)
   \\ = & 
   \sum_{ i}\int d\mathbf{r} e^{-i \mathbf{k} \cdot \mathbf{r}  -i n (-\mathbf{g}_1 + \mathbf{g}_2) \cdot \mathbf{r}   } \mathbf{A} \cdot \mathbf{r} \left( e^{i \mathbf{k} \cdot \mathbf{r}_i}u^{i, s}_{\alpha, \mathbf{k}} \phi_{\alpha}(\mathbf{r} - \mathbf{r}_i)  \right).
\end{split}
\end{equation}
\end{widetext}
The above manipulation involves taking $\mathbf{r} \rightarrow \mathbf{r} - \mathbf{R}$, and using $e^{i \mathbf{G} \cdot \mathbf{R}} = 1$.  Additionally, the overlap integral of the free particle and the initial state are assumed orthogonal, so the term proportional to $\mathbf{R}$ vanishes \cite{Schler2020}.

\begin{figure*}[bt!]
\includegraphics[width=2\columnwidth]{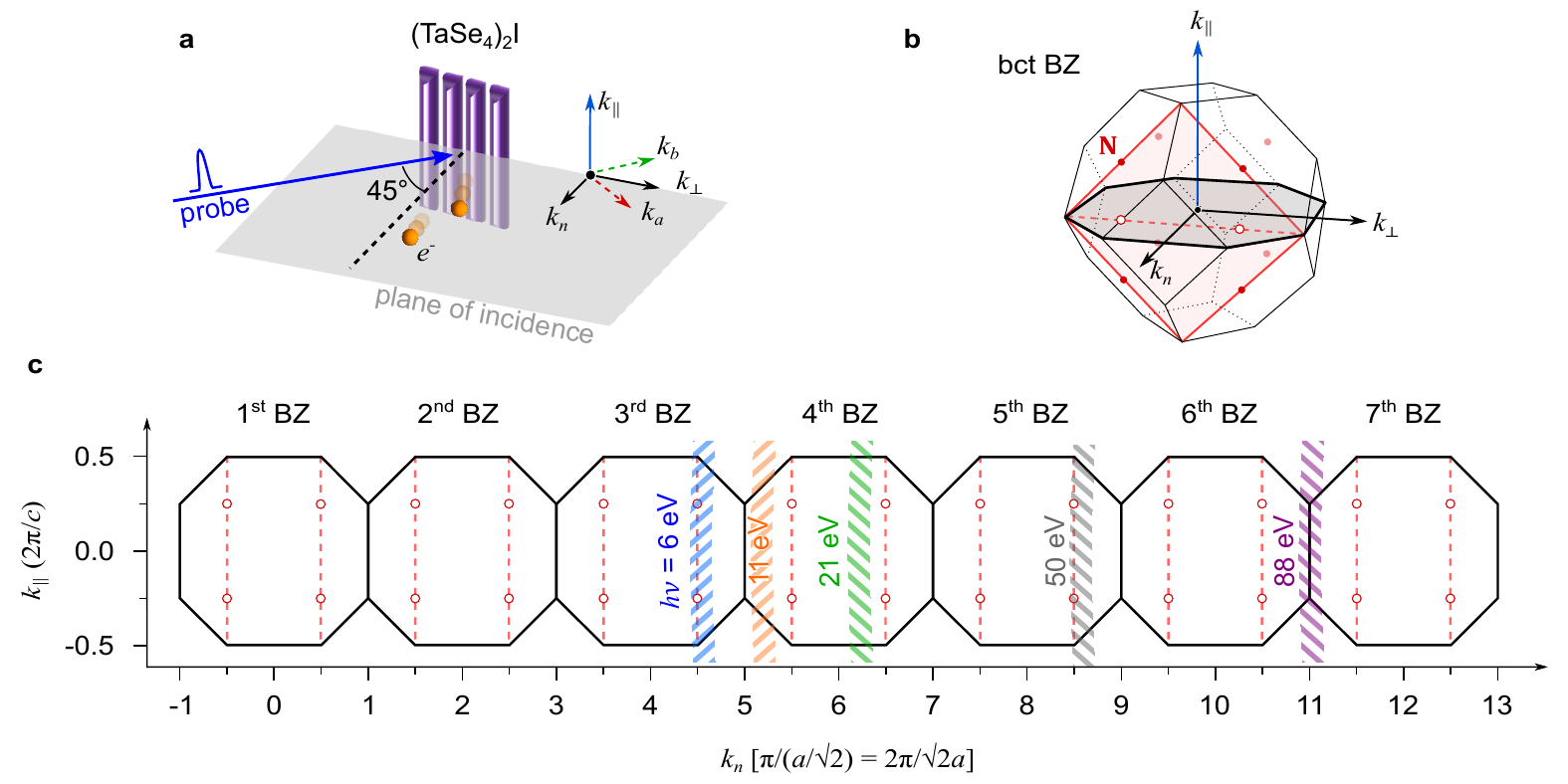}
\caption{\textbf{Estimation of the out-of-plane momentum $k_n$} \textbf{(a)} Sample and photoemitting beam geometry for this work. $k_{\parallel}$ is parallel to the chain. $k_{\perp}$ is perpendicular to the chain direction, and $k_n$ is normal to the natural cleavage plane (110). \textbf{(b)} 3D body-centered-tetragonal Brillouin zone (bct BZ) of (TaSe$_4$)$_2$I. A 2D projection on $k_{\perp}$-$k_\text{n}$ plane is shown in the gray shaded area and the $k_{\perp}$-$k_{\parallel}$ plane containing the $N$-points in the red shaded area. For the latter, only one of the two planes are shown for simplicity. \textbf{(c)} Extended BZ in the $k_\text{n}$ direction. The shaded area denotes the estimated values of $k_\text{n}$ for $h\nu$ = 6 eV and selected photon energies are shown.}\label{FigS_innerpot}
\end{figure*}

We can tease out an analytic expression from this statement.  First, we let $\mathbf{r}^{\prime} = \mathbf{r} - \mathbf{r}_i$.  This gives:
\begin{equation}
\label{analyticMs1}
\resizebox{1\hsize}{!}{$
\begin{aligned}
    &M_s(\mathbf{k}, n, \mathbf{A})
    \\
    &= \sum_i \int d \mathbf{r}^{\prime} e^{-i (\mathbf{k} -n (-\mathbf{g}_1 + \mathbf{g}_2) ) \cdot \mathbf{r}^{\prime} }e^{-i n \mathbf{G} \cdot \mathbf{r}_i} \mathbf{A} \cdot (\mathbf{r}^{\prime} + \mathbf{r}_i ) u^{i, s}_{\alpha, \mathbf{k}} \phi_{\alpha}(\mathbf{r}^{\prime})
    \\&=
    \sum_i  e^{-i n (-\mathbf{g}_1 + \mathbf{g}_2) \cdot \mathbf{r}_i} u^{i, s}_{\alpha, \mathbf{k}} \mathbf{A} \cdot \left[ \int d \mathbf{r}^{\prime} e^{-i \Tilde{\mathbf{k}} \cdot \mathbf{r}^{\prime} }  (\mathbf{r}^{\prime} + \mathbf{r}_i )  \phi_{\alpha}(\mathbf{r}^{\prime}) \right],
\end{aligned}
$}
\end{equation}
where $\Tilde{\mathbf{k}} \equiv \mathbf{k} + n (-\mathbf{g}_1 + \mathbf{g}_2)$.  We now examine the integral that goes as $f(\Tilde{\mathbf{k}}) = \int d \mathbf{r}^{\prime} e^{-i \Tilde{\mathbf{k}} \cdot \mathbf{r}^{\prime}} \phi_{\alpha}(\mathbf{r}^{\prime})$.  Since the orbital function, $\phi_{\alpha}(\mathbf{r}^{\prime})$, takes the form of $e^{- |\mathbf{r}^{\prime}| / a_0}$, then this is just a Fourier transform of an s-wave orbital.  It becomes
\begin{equation}
    f(\Tilde{\mathbf{k}}) = \frac{8 \pi }{a_0} \frac{1}{(\frac{1}{a_0^2} + |\Tilde{\mathbf{k}}|^2)^2}.
\end{equation}
We can employ a Feynman trick to further simplify Eq.~\eqref{analyticMs1} such that
\begin{equation}
\begin{split}
    \int d \mathbf{r}^{\prime} e^{-i \Tilde{\mathbf{k}} \cdot \mathbf{r}^{\prime}} \mathbf{r}^{\prime} \phi_{\alpha}(\mathbf{r}^{\prime}) =& i\nabla_{\mathbf{\tilde{k}}} f( \Tilde{\mathbf{k}}) 
    \\
    =& \frac{- 32 \pi i \Tilde{\mathbf{k}}}{a_0} \frac{1}{(\frac{1}{a_0^2} + |\Tilde{\mathbf{k}}|^2)^3}.
\end{split}
\end{equation}
Using these two integrals, we can rewrite Equation \ref{analyticMs1} as 
\begin{widetext}
\begin{equation}
    M_s(\mathbf{k}, n , \mathbf{A}) = \frac{8 \pi}{a_0}\sum_i e^{-i n (-\mathbf{g}_1 + \mathbf{g}_2) \cdot \mathbf{r}_i} u^{i, s}_{\alpha, \mathbf{k}} \mathbf{A} \cdot \left[ \frac{ - 4  i \Tilde{\mathbf{k}} }{(\frac{1}{a_0^2} + |\Tilde{\mathbf{k}}|^2)^3} + \frac{\mathbf{r}_i}{(\frac{1}{a_0^2} + |\Tilde{\mathbf{k}}|^2)^2}  \right].
\end{equation}
\end{widetext}
The normalized helicity-dependent ARPES signal is then given by averaging $|M_s|^2$ over the spin of the final state.

\begin{figure}[bt!]
\includegraphics[width=1\columnwidth]{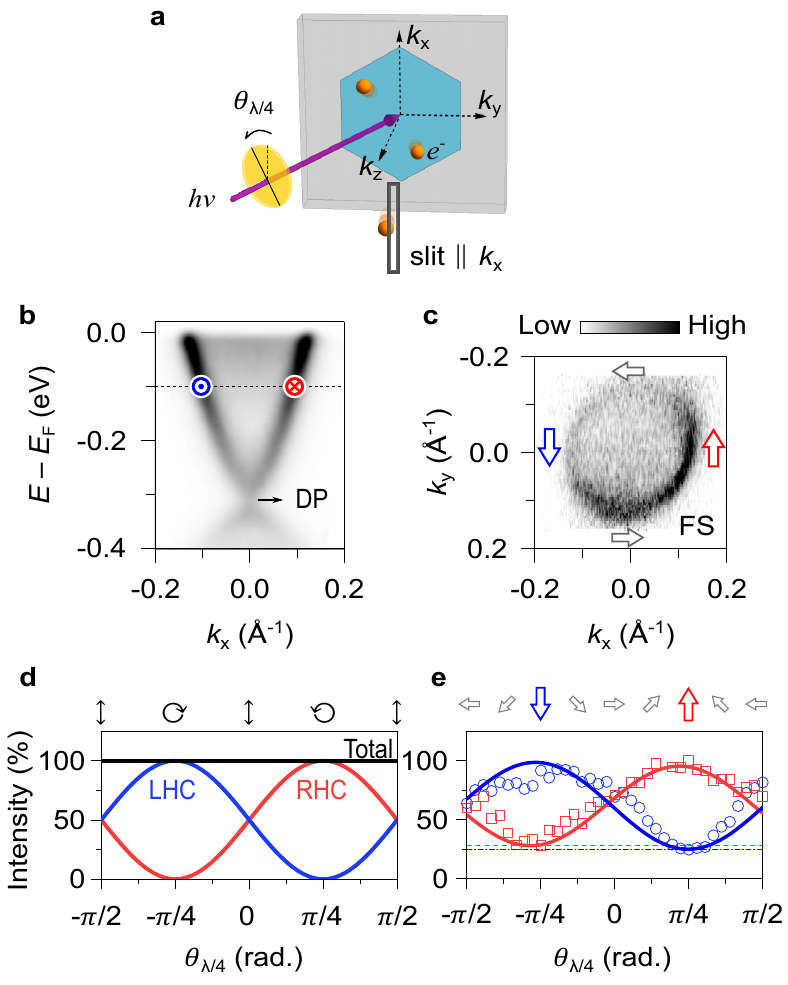}
\caption{Helicity-dependent ARPES spectra of Bi$_2$Se$_3$. (a) Measurement geometry of the ARPES setup. (b) Energy (E) relative to the Fermi level (E$_f$) versus momentum along the k$_x$ direction measured at $T$ = 7 K. DP denoted the Dirac point. (c) Constant energy cut taken at Fermi surface. The ideal tangential spin texture of the topological Dirac surface states is indicated with arrows. (d) Components of the left- and right-handed circularly polarization (LHC \& RHC) of probe beam at $\theta_{\lambda/4}$, the angle between the $\lambda$/4 waveplate fast axis and the analyzer slit. (e) $\theta_{\lambda/4}$ dependence of the photoemission intensities for the Dirac surface state at $E-E_F$ = -0.1 eV. The integrated regions are marked with symbols in (b). For each region, the experimental data (symbols) are fitted with a helicity-independent background (dashed/dash-dotted lines) and a cos$^2\theta$ function.}\label{FigS_BiSe}
\end{figure}

\subsection{IV. Estimation of the out-of-plane momentum $k_n$}

As is well known, the in-plane momenta (referred to in this work as $k_{\perp}$ and $k_{\parallel}$) are conserved in the photoemission process in an ARPES experiment \cite{Sobota2021}. The out-of-plane momentum ($k_n$) is not conserved but can be estimated using equation below:
\begin{equation}
    k_n = \sqrt{\frac{2m}{\hbar^2}[(h\nu-W)\text{cos}^2\theta+V_0]},
\end{equation}
where $\theta$ is the emission angle of detected photoelectrons, $W$ the workfunction, and $V_0$ the inner potential that can be obtained using a tunable light source \cite{Sobota2021}. Following previous photon-energy dependent ARPES works on (TaSe$_4$)$_2$I (Ref.\cite{TournierColletta2013}), we used $V_0$ = 16 eV and $W \sim$ 4.5 eV to estimate the range for $k_n$ in our 6 eV laser-ARPES setup. Fig.~\ref{FigS_innerpot}(c) shows this range for $k_n$ in the extended BZ. The corresponding $k_{\text{n}}$ for $h\nu$ = 6 eV is weighted towards the center of BZ, but its $k_{\perp}$-$k_{\parallel}$ planes are close to those containing N-points, i.e. the red shaded region in panel (b). The accessible $k_n$ ranges for various photon energies that are commonly used for laser-ARPES setups and 88 eV used in Ref.\cite{TournierColletta2013} are also shown.

\subsection{V. Helicity-dependent laser ARPES on Bi$_2$Se$_3$}

Here we present helicity-dependent laser ARPES results on the prototypical topological insulator (TI) Bi$_2$Se$_3$. The sample was cleaved inside the UHV chamber at pressure below $1\times 10^{-10}$ mbar at 7 K. The measurement geometry is the same as that for the (TaSe$_4$)$_2$I and the sample was mounted in (111) direction as shown in Fig. \ref{FigS_BiSe}(a). The Dirac point is located at $E-E_F = -0.3$ eV (labelled DP in Fig.\ref{FigS_BiSe}(b)). In contrast to the characteristic radial/hedgehog spin texture near a KW point in a chiral crystal (see main text and Refs.~\cite{Han2012, Gatti2020, Sakano2020}), the topological surface bands of a TI has a tangential spin texture near the Dirac point as studied both experimentally and theoretically in Refs.\cite{CPark2012, gedik2011, xidaiDirac}. This is illustrated for Bi$_2$Se$_3$ with arrows in the constant energy map in Fig.~\ref{FigS_BiSe}(c). Constrained by spin-momentum locking,  opposite spins reside on $k=\pm k_0$ at constant energy, as marked in $E-E_F = -0.1$ eV of Fig. \ref{FigS_BiSe}(b).

Figure \ref{FigS_BiSe}(e) presents the $\theta_{\lambda/4}$-dependent photoemission intensity on each side of the Dirac surface bands (Fig.~\ref{FigS_BiSe}(b)). In our measurement geometry, the $\theta_{\lambda/4}$ indicates the angle between the fast-axis of the quarter waveplate and the analyzer slit. As plotted in Fig.~\ref{FigS_BiSe}(d), while the modulation of the $\theta_{\lambda/4}$ keeps the total intensity the same (as that of a linearly polarized input beam), the light helicity is controlled by changing the relative intensities between the left- and right-handed circularly polarization (LHC \& RHC). The generated probe beam right after the waveplate is purely circular (either LHC or RHC) at $\theta_{\lambda/4} = \pm \pi / 4$.

Similar to previous results, the photoemission intensities of Bi$_2$Se$_3$ surface states with opposite spins follow the intensities of LHC/RHC components, showing a maximum and a minimum intensities at $\theta_{\lambda/4} = \pm \pi / 4$, respectively (Fig.~\ref{FigS_BiSe}(e)). To obtain the peak positions better, each of the data is fitted with a $\cos^2 \theta $ function, similar to LHC/RHC intensities, with a constant offset. 

The broadening of the peaks in Fig.~\ref{FigS_BiSe}(e) can be understood as the warping effect of the orbital angular momentum that gives additional pseudospin texture.  This warping results in multiple components in the $\theta_{\lambda/4}$-dependent intensity. Each of the peaks are best fit with three components, which may be closely relevant to the $C_\text{3}$ symmetry of Bi$_2$Se$_3$ and Bi$_2$Te$_3$ \cite{gedik2011, Fu2009, Jung2011, Mirhosseini2012}. The warping effects are conventionally shown with the constant energy contours of CD showing sign reversals around the surface bands, or the intensity plots as a function of the azimuth angle from the Dirac point, where $\sin \theta$, $\sin 3\theta$, and $\sin 6\theta$ were demonstrated (e.g. see Fig.~3 in Ref.\cite{Jung2011}). The major difference of Fig. \ref{FigS_BiSe}(e) from previous work is that we fix the momentum/energy position and change the probe helicity ($\theta_{\lambda/4}$). In comparison, previous works fix the helicity and change the momentum within the band.
As described in the main text, the $\theta_{\lambda/4}$-dependence of (TaSe$_4$)$_2$I with an asymmetric radial spin texture near the N KW point is distinct from the case of Bi$_2$Se$_3$ which has a symmetric tangential spin texture. As shown here, the photoemission intensity is symmetric across the Dirac point in Bi$_2$Se$_3$, in contrast to (TaSe$_4$)$_2$I which shows a asymmetry across the N KW point (main text).

\end{document}